\documentclass{PoS}
\usepackage{amsmath}
\usepackage{cancel}

\newcommand{\al}{\ensuremath{\alpha} }
\newcommand{\be}{\ensuremath{\beta} }
\newcommand{\ga}{\ensuremath{\gamma} }
\newcommand{\De}{\ensuremath{\Delta} }
\newcommand{\lsim}{\ensuremath{\lesssim} }

\newcommand{\X}{\ensuremath{\!\times\!} }
\newcommand{\Sb}{\ensuremath{\cancel{S^4}} }
\newcommand{\MSbar}{\ensuremath{\overline{\textrm{MS}}} }
\newcommand{\refcite}[1]{Ref.~\cite{#1}}
\newcommand{\fig}[1]{Fig.~\ref{#1}}
\newcommand{\secref}[1]{Section~\ref{#1}}

\title{MCRG study of 8 and 12 fundamental flavors}
\ShortTitle{MCRG study of 8 and 12 fundamental flavors}

\author{\speaker{Gregory Petropoulos}, Anqi Cheng, Anna Hasenfratz and David Schaich \\
  Department of Physics, University of Colorado, Boulder, CO 80309 \\
  Email: \email{gregory.petropoulos@colorado.edu}
}

\abstract{ 
  We study the renormalization group properties of SU(3) gauge theories with $N_f = 8$ and 12 nearly-massless fermions, using Monte Carlo Renormalization Group (MCRG) two-lattice matching techniques to predict bare step-scaling functions $s_b$.
  Traditional MCRG two-lattice matching requires that the renormalization scheme be optimized for each bare lattice coupling, so that $s_b$ is a composite of many different discrete \be functions.
  We propose an improved procedure that uses the Wilson flow to eliminate the need for this optimization of the RG blocking transformation.
  While our 12-flavor results indicate an infrared fixed point, $s_b$ for $N_f = 8$ is significantly different from zero until strong-coupling lattice artifacts obstruct two-lattice matching.
  Although both procedures produce qualitatively similar bare step-scaling functions, the new $s_b$ obtained by combining the Wilson flow with MCRG two-lattice matching have the distinct advantage of corresponding to unique discrete \be functions.
}

\FullConference{The 30th International Symposium on Lattice Field Theory \\ June 24--29, 2012 \\ Cairns, Australia}

\begin{document}
\section{Introduction and overview of methods \& results} 
In recent years, many groups have initiated lattice investigations of strongly-coupled gauge--fermion systems beyond QCD.
While the ultimate goal of these efforts is to explore potential new physics beyond the standard model, an essential step is to improve our theoretical understanding of the basic properties of these non-perturbative systems.
Here we study the renormalization group properties of SU(3) gauge theories with $N_f = 8$ and 12 nearly-massless fermions in the fundamental representation, through the Monte Carlo Renormalization Group (MCRG) two-lattice matching technique.
This is one of several complementary analyses we are currently carrying out, two more of which (investigating Dirac eigenmode scaling and finite-temperature transitions) are discussed in other contributions to these proceedings~\cite{Hasenfratz:2012fp, Schaich:2012fr}.
Recent references on SU(3) gauge theories with $N_f = 8$ and 12 include~\cite{Fodor:2012uw, Fodor:2012et, Aoki:2012eq, Deuzeman:2012ee, Lin:2012iw}; earlier works are reviewed in \refcite{Giedt:2012LAT}.

In Refs.~\cite{Hasenfratz:2011xn, Hasenfratz:2011np}, one of us studied MCRG two-lattice matching for the 12-flavor system with nHYP-smeared staggered actions very similar to those we use here.
Our gauge action includes both fundamental and adjoint plaquette terms, with coefficients related by $\be_A = -0.25\be_F$.
The negative adjoint plaquette term lets us avoid a well-known spurious ultraviolet fixed point caused by lattice artifacts, and implies $\be_F = 12 / g^2$ at the perturbative level.
In our fermion action, we use nHYP smearing with parameters $(0.5, 0.5, 0.4)$, instead of the $(0.75, 0.6, 0.3)$ used by Refs.~\cite{Hasenfratz:2011xn, Hasenfratz:2011np}.
By changing the nHYP-smearing parameters in this way, we can access stronger couplings without encountering numerical problems.
At such strong couplings, for both $N_f = 8$ and $N_f = 12$ we observe a lattice phase in which the single-site shift symmetry (``$S^4$'') of the staggered action is spontaneously broken (``$\Sb$'')~\cite{Cheng:2011ic, Schaich:2012fr}.\footnote{\refcite{Deuzeman:2012ee} recently interpreted the \Sb lattice phase in terms of relevant next-to-nearest neighbor interactions.}
In this work we only investigate couplings weak enough to avoid the \Sb lattice phase.

In the next section, we review how the MCRG two-lattice matching technique determines the step-scaling function $s_b$ in the bare parameter space.
Although working entirely with bare parameters would be disadvantageous if our aim were to produce renormalized phenomenological predictions for comparison with experiment, our current explorations of the phase structures of the 8- and 12-flavor systems benefit from this fully non-perturbative RG approach, especially for relatively strong couplings.
In \secref{sec:MCRGresults} we present our results from the traditional MCRG two-lattice matching technique.
While our 8-flavor $s_b$ is significantly different from zero, for $N_f = 12$ we observe $s_b \lsim 0$ for $\be_F < 8$, indicating an infrared fixed point (IRFP).

We emphasize that while the existence of an IRFP is physical (scheme-independent), the coupling at which it is located depends on the choice of renormalization scheme.
A limitation of traditional MCRG two-lattice matching is the need to optimize the RG blocking transformation separately for each lattice coupling $\be_F$.
As we explain below, this optimization forces us to probe a different renormalization scheme for each $\be_F$, so that the bare step-scaling function we obtain is a composite of many different discrete \be functions.

To address this issue, in \secref{sec:WMCRG} we propose a new, improved procedure that predicts a bare step-scaling function corresponding to a unique \be function.
This improved procedure applies the Wilson flow~\cite{Narayanan:2006rf, Luscher:2010iy} to the lattice system before performing the RG blocking transformation.
Because the Wilson flow moves the system in the infinite-dimensional space of lattice-action terms without changing the lattice scale, we can use it to approach the renormalized trajectory corresponding to a fixed RG blocking transformation.
By optimizing the flow time $t_f$ at each coupling, all with the same renormalization scheme, we can carry out the two-lattice matching without a need for further optimization.
We present some promising but preliminary results of this approach in \secref{sec:WMCRGresults}.

\section{Two-lattice matching procedures and the need for optimization} 
Two-lattice matching is most easily described in the context of confining systems, where it locates pairs of couplings $(\be_F, \be_F')$ for which lattice correlation lengths obey $\xi(\be_F) = 2\xi(\be_F')$.
We proceed by repeatedly applying RG blocking transformations (with scale factor $s = 2$) to lattices of volume $24^3\X48$, $12^3\X24$ and $6^3\X12$.\footnote{We are currently generating larger lattices up to $32^3\X64$, which will permit additional consistency checks.}
Under RG blocking on the $m = 0$ critical surface, the system flows toward the renormalized trajectory in irrelevant directions, and along it in the relevant direction.
By blocking the larger lattices (with $\be_F$) $n_b$ times and the smaller lattices (with $\be_F'$) only $n_b - 1$ times, we obtain blocked systems with the same lattice volume.
If these blocked systems have both flowed to the same point on the renormalized trajectory, then we can conclude that $\xi(\be_F) = 2\xi(\be_F')$ on the unblocked systems, as desired.

We determine whether the blocked systems have flowed to the same point on the renormalized trajectory by matching several short-range gauge observables: the plaquette, all three six-link loops, and two planar eight-link loops.
For a given $\be_F$, each observable may predict a different $\De\be_F \equiv \be_F - \be_F'$.
The spread in these results is a systematic error that dominates our uncertainties.

In an IR-conformal system, the gauge coupling that is relevant at the perturbative gaussian FP becomes irrelevant at the IRFP.
The renormalized trajectory connects these two fixed points.
When RG flows approach this renormalized trajectory, the two-lattice matching can be performed and interpreted the same way as in confining systems.
In this region the gauge coupling flows to stronger couplings, $\De\be_F > 0$ corresponding to a negative RG \be function.
The situation is less clear at stronger couplings where we might na\"ively expect backward flow.
If there is no ultraviolet FP in this region to drive the RG flow along a renormalized trajectory, the two-lattice matching might become meaningless.
This issue affects every method that attempts to determine the flow of the gauge coupling in IR-conformal systems at strong coupling.
In all published studies that report an IRFP, backward flow has only been observed in a very limited range of couplings in the immediate vicinity of the IRFP (cf.~\refcite{Giedt:2012LAT}).

Since we can block our lattices only a few times, we must optimize the two-lattice matching by requiring that consecutive RG blocking steps yield the same $\De\be_F$.
We identify the optimized $\De\be_F$ with the bare step-scaling function $s_b$.
In the following subsections, we describe two different ways to perform this optimization.
The traditional technique optimizes the RG blocking transformation (renormalization scheme).
The new method we propose in \secref{sec:WMCRG} instead applies the Wilson flow to the lattice system prior to RG blocking, and optimizes the flow time $t_f$.


\subsection{Traditional MCRG} 
\begin{figure}[htpb]
  \centering
  \includegraphics[width=0.45\linewidth]{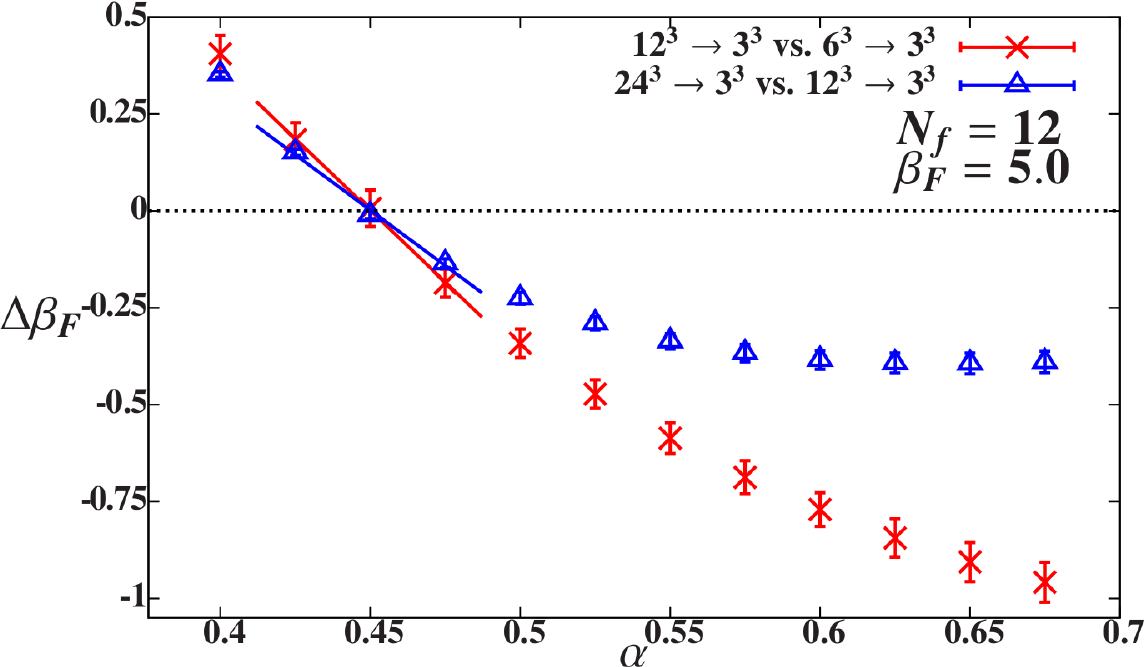}\hfill
  \includegraphics[width=0.45\linewidth]{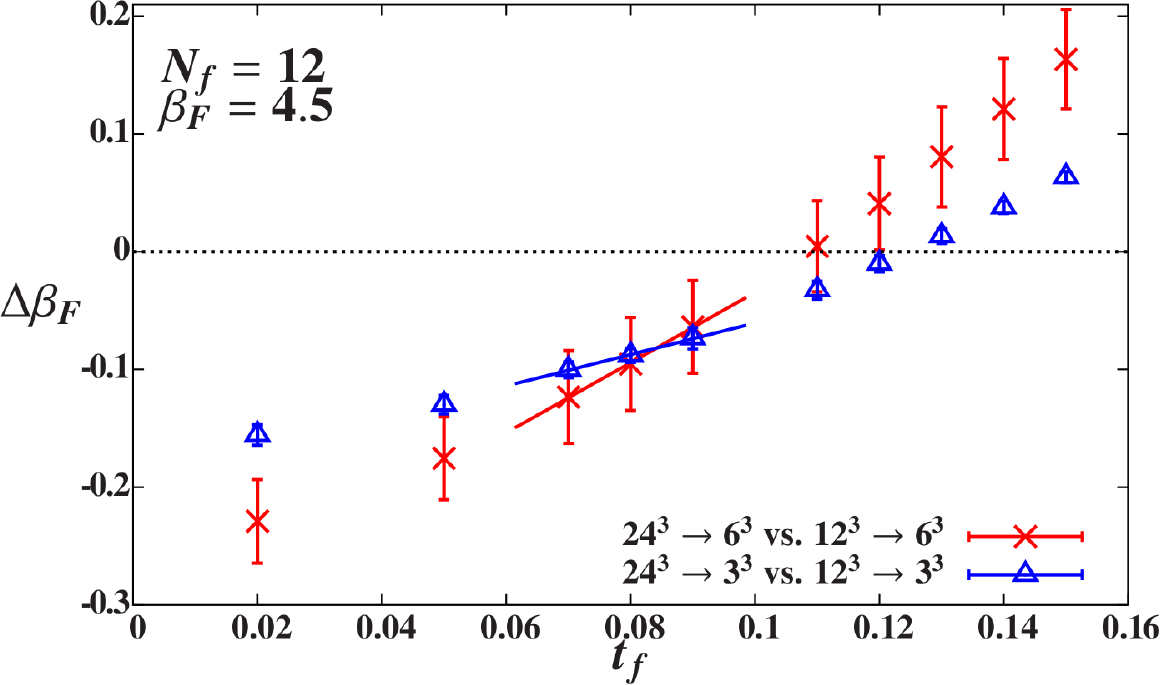}
  \caption{Examples of two-lattice matching optimization for 12-flavor systems.  Left: Optimization of the HYP-smearing parameter \al in the RG blocking transformation, for $\be_F = 5.0$.  Right: Optimization of the Wilson flow time $t_f$ with fixed $\al = 0.5$, for $\be_F = 4.5$.  In both cases, the uncertainties on the data points are dominated by averaging over the different observables as described in the text.}
  \label{fig:opt}
\end{figure}

As in Refs.~\cite{Hasenfratz:2011xn, Hasenfratz:2011np}, we use RG blocking transformations that include two sequential HYP smearings with parameters $(\al, 0.2, 0.2)$, and optimize \al as shown in the left panel of \fig{fig:opt}.
Qualitatively, this optimization finds the renormalization scheme for which the renormalized trajectory passes as close as possible to the lattice system with coupling $\be_F$.
Without optimization, residual flows in irrelevant directions can distort the results: this is the reason $\De\be_F$ changes with \al in \fig{fig:opt}, and also explains why increasing the number of blocking steps reduces this $\al$-dependence.

The downside of optimizing the RG blocking transformation in this manner is that we have to use a different renormalization scheme for each $\be_F$.
As a result, the bare step-scaling function we obtain is a composite of many different discrete \be functions.

\subsection{\label{sec:WMCRG}Wilson-flowed MCRG} 
As an alternative to optimizing the RG blocking transformation, and thus changing the renormalization scheme at each coupling $\be_F$, here we propose to use the Wilson flow to move the lattice system as close as possible to the renormalized trajectory of a fixed renormalization scheme.

The Wilson flow is a continuous smearing transformation~\cite{Narayanan:2006rf} that can be related to the \MSbar running coupling in perturbation theory~\cite{Luscher:2010iy}.
Refs.~\cite{Fodor:2012td, Fodor:2012qh} recently used the Wilson flow to compute a renormalized step-scaling function in a way similar to Schr\"odinger functional methods.
While this approach appears very promising, it is based on perturbative relations that are only fully reliable at weak coupling.
Here we do not use this perturbative connection, instead applying the Wilson flow as a continuous smearing that removes UV fluctuations.
The Wilson flow moves the system along a surface of constant lattice scale in the infinite-dimensional action-space; it is not a renormalization group transformation and does not change the IR properties of the system.

Our goal is to use a one-parameter Wilson flow transformation to move the lattice system as close as possible to the renormalized trajectory of our fixed RG blocking transformation.
We proceed by carrying out two-lattice matching after applying the Wilson flow for a flow time $t_f$ on all lattice volumes.
(The Wilson flow is run only on the unblocked lattices, not in between RG blocking steps.)
As above, since we can block our lattices only a few times, we must optimize $t_f$ by requiring that consecutive RG blocking steps yield the same $\De\be_F$, as shown in the right panel of \fig{fig:opt}.
As for traditional MCRG, increasing the number of blocking steps reduces the dependence on the optimization parameter; in the limit $n_b \to \infty$, our results would be independent of $t_f$.

With Wilson-flowed MCRG we can efficiently determine bare step-scaling functions that correspond to unique RG \be functions.
This new capability opens up interesting directions for future studies.
By comparing different \be functions around the perturbative gaussian FP, we can study scaling violations in the lattice system.
In IR-conformal systems, we can investigate the scheme-dependence of the \be function near the IRFP, an issue explored in perturbation theory by \refcite{Ryttov:2012nt}.

\section{Results} 
\subsection{\label{sec:MCRGresults}Traditional MCRG} 
\begin{figure}[htpb]
  \includegraphics[width=0.45\linewidth]{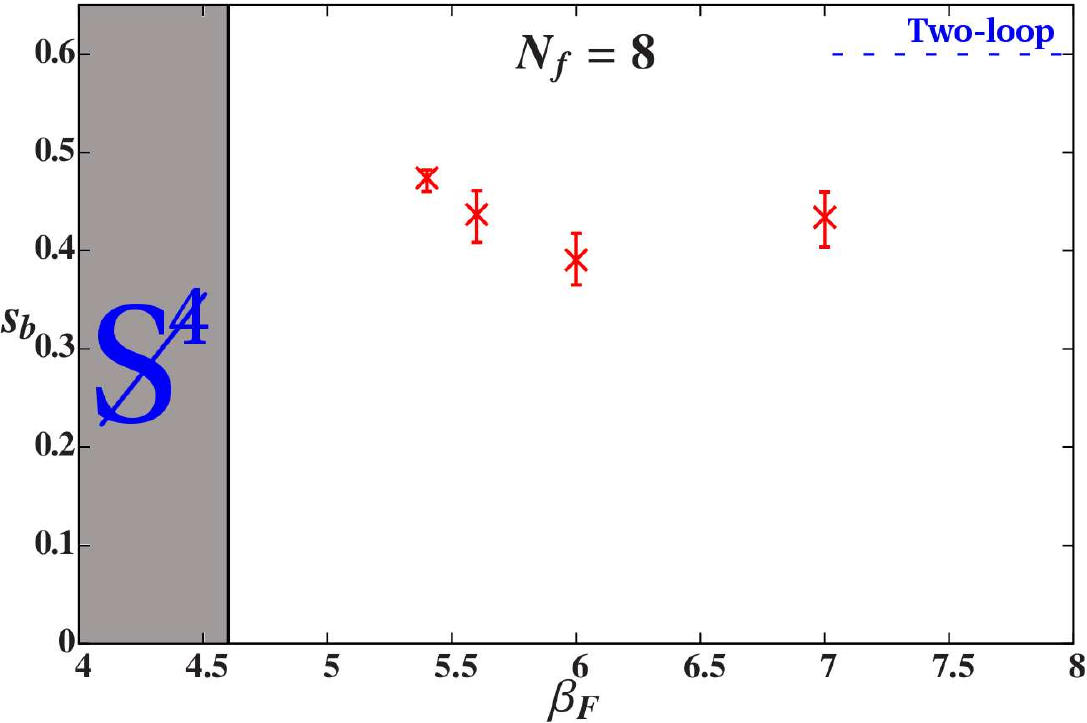}\hfill
  \includegraphics[width=0.45\linewidth]{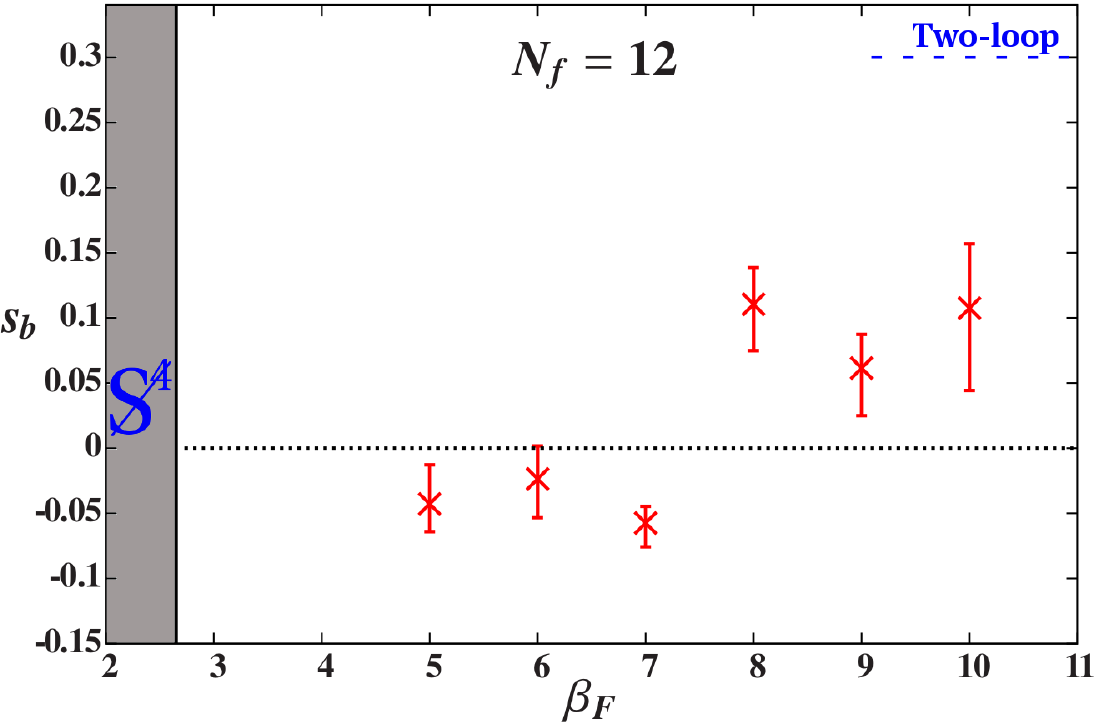}
  \caption{Results for the bare step-scaling function $s_b$ from traditional MCRG two-lattice matching with $24^3\X48$, $12^3\X24$ and $6^3\X12$ lattice volumes, for $N_f = 8$ (left) and $N_f = 12$ (right).  The blue dashed lines are perturbative predictions for asymptotically weak coupling.}
  \label{fig:MCRG}
\end{figure}

Our results for the bare step-scaling function $s_b$ from traditional MCRG two-lattice matching are shown in \fig{fig:MCRG}.
On the largest $24^3\X48$ lattices that we use in this current study, we work with fermion masses $m = 0.0025$ to stay near the $m = 0$ critical surface.
Under RG blocking with scale factor $s$, the fermion mass changes as $s^{1 + \ga_m}$ where $\ga_m$ is the mass anomalous dimension.
Therefore we use $m = 0.01$ on $12^3\X24$ and $m = 0.02$ on $6^3\X12$ lattices.
We have explicitly checked that these masses are small enough to introduce only negligible finite-mass effects, by generating lattices with $m = 0$ for some points and obtaining indistinguishable results.

While our 8-flavor results for $s_b$ are significantly different from zero for all couplings we can explore, for $N_f = 12$ we find $s_b \lsim 0$ for $\be_F < 8$, indicating an IRFP.
Recall that our optimization of the RG blocking transformation means that we use a different renormalization scheme for each coupling $\be_F$, so these bare step scaling functions are composites of several different discrete \be functions.
For example, with $N_f = 12$ at $5 \leq \be_F \leq 6$, our optimization selects renormalization schemes with the fixed point near $\be_F$, so that $s_b$ is roughly consistent with zero over an extended range.
Both our $N_f = 8$ and 12 simulations encounter the \Sb lattice phase at strong coupling, where we cannot perform matching.
As in Refs.~\cite{Hasenfratz:2011xn, Hasenfratz:2011np}, we do not explore weak enough couplings to recover the two-loop perturbative predictions $s_b \approx 0.6$ (0.3) for $N_f = 8$ (12).

As mentioned above, the error bars shown in \fig{fig:MCRG} are dominated by the spread in results from matching four-, six- and eight-link loops.
Each of these observables can predict a different optimal $\al$, and for fixed \al each can predict a different $\De\be_F$.
Preliminary results presented at the conference determined uncertainties from the full spread of optimal \al predicted by the different observables.
Here, instead, we average $\De\be_F$ for the different observables at fixed $\al$, and use these combined data to optimize \al and find the associated uncertainties.

\subsection{\label{sec:WMCRGresults}Wilson-flowed MCRG} 
\begin{figure}[htpb]
  \includegraphics[width=0.45\linewidth]{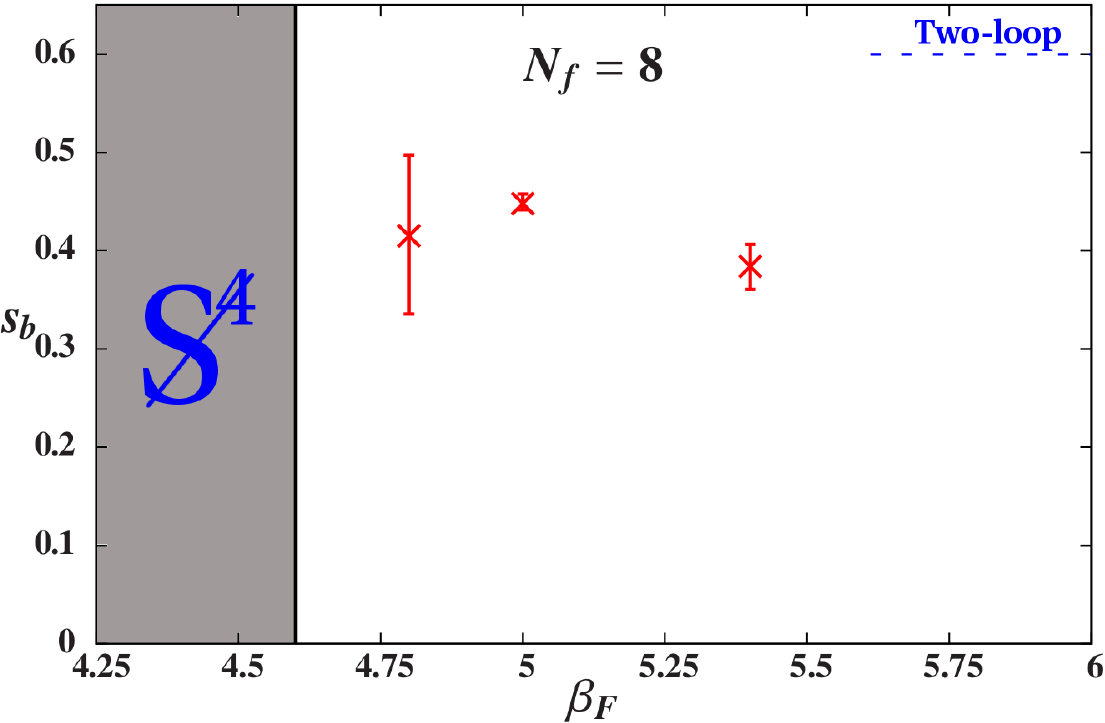}\hfill
  \includegraphics[width=0.45\linewidth]{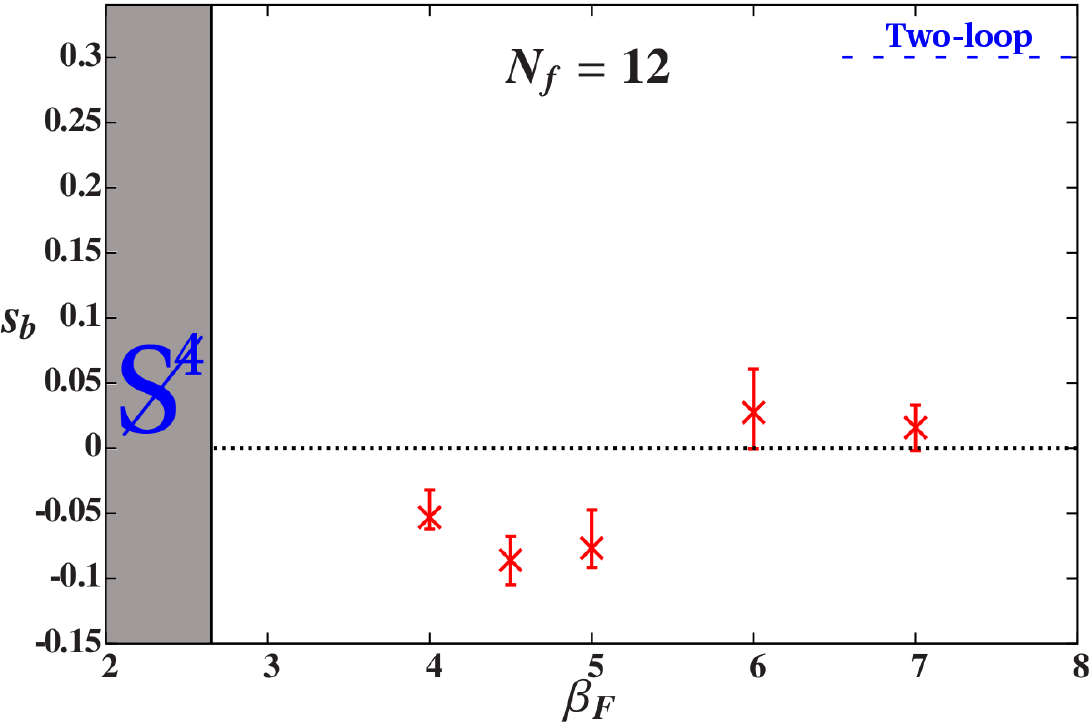}
  \caption{Preliminary results for the bare step-scaling function $s_b$ from Wilson-flowed MCRG two-lattice matching with $24^3\X48$ and $12^3\X24$ lattice volumes, for $N_f = 8$ (left) and $N_f = 12$ (right) with fixed HYP-smearing parameters $(0.5, 0.2, 0.2)$.  The blue dashed lines are perturbative predictions for asymptotically weak coupling.}
  \label{fig:WMCRG}
\end{figure}

\fig{fig:WMCRG} presents our results for the bare step-scaling function $s_b$ from Wilson-flowed MCRG two-lattice matching.
We continue using two sequential HYP smearings in our RG blocking transformation, but now fix the smearing parameters to $(0.5, 0.2, 0.2)$.
We again use $24^3\X48$ and $12^3\X24$ lattices with fermion masses $m = 0.0025$ and $m = 0.01$, respectively, and determine uncertainties in the same way as described in the previous subsection.
In this preliminary study we don't yet employ the volume-corrected optimization discussed in \refcite{Hasenfratz:2011xn}; for our lattice volumes, \refcite{Hasenfratz:2011xn} found that neglecting this finite-volume correction introduces only a small additional error.
Our final results will use the appropriate optimization, and will also present further consistency checks from larger lattices up to $32^3\X64$.

While our results in \fig{fig:WMCRG} from combining the Wilson flow with MCRG two-lattice matching are qualitatively similar to the results of the traditional approach in \fig{fig:MCRG}, we can now identify the bare step-scaling function $s_b$ with a unique discrete \be function.
In the renormalization scheme defined by our RG blocking transformation with HYP-smearing parameters $(0.5, 0.2, 0.2)$, we find a 12-flavor IRFP at $5 < \be_F^{\star} \lsim 6$.
Although $\be_F^{\star}$ is scheme-dependent, the existence of this IRFP is physical.
We are currently exploring other choices of renormalization schemes, to non-perturbatively investigate the scheme-dependence of the \be function near the IRFP~\cite{Ryttov:2012nt}.
At weaker couplings, we will also attempt to use similar explorations to study scaling violations in our lattice systems.

\section{Conclusions} 
We have proposed a new, improved MCRG two-lattice matching procedure that uses the Wilson flow to eliminate the need for optimization of the RG blocking transformation.
Both traditional MCRG and Wilson-flowed MCRG produce bare step scaling functions $s_b$ that indicate an infrared fixed point for SU(3) gauge theory with $N_f = 12$ fundamental fermions, while $s_b$ for $N_f = 8$ is significantly different from zero in the accessible range of lattice couplings.
The results obtained by combining the Wilson flow with two-lattice matching correspond to a unique \be function, unlike $s_b$ from traditional MCRG, which is a composite of many different discrete \be functions.

\section*{Acknowledgments} 
We thank D\'aniel N\'ogr\'adi for helpful comments on the Wilson flow.
This research was partially supported by the U.S.~Department of Energy (DOE) through Grant No.~DE-FG02-04ER41290 (A.~C., A.~H.\ and D.~S.), and by the DOE Office of Science Graduate Fellowship (SCGF) Program made possible by the American Recovery and Reinvestment Act of 2009 and administered by the Oak Ridge Institute for Science and Education managed by Oak Ridge Associated Universities under Contract No.~DE-AC05-06OR23100.
Our code is based in part on the MILC Collaboration's public lattice gauge theory software,\footnote{\texttt{http://www.physics.utah.edu/$\sim$detar/milc/}} and on the code distributed with \refcite{Borsanyi:2012zs}.
Numerical calculations were carried out on the HEP-TH and Janus clusters at the University of Colorado, the latter supported by National Science Foundation (NSF) Grant No.~CNS-0821794; at Fermilab under the auspices of USQCD supported by the DOE SciDAC program; and at the San Diego Computing Center through XSEDE supported by NSF Grant No.~OCI-1053575.

\bibliographystyle{utphys}
\bibliography{pos2012}
\end{document}